\documentclass[pre,preprint,notitlepage]{revtex4-2}%
\usepackage{amssymb}
\usepackage{amsfonts}
\usepackage{amsmath}
\usepackage{ulem}
\usepackage{xcolor}
\usepackage{graphicx}%
\providecommand{\U}[1]{\protect\rule{.1in}{.1in}}

\begin{document}
\title{A microscopic approach to crystallization: challenging the classical/non-classical dichotomy}
\author{James F. Lutsko}
\homepage{http://www.lutsko.com}
\email{jlutsko@ulb.be}
\author{C\'{e}dric Schoonen}
\affiliation{Center for Nonlinear Phenomena and Complex Systems CP 231, Universit\'{e}
  Libre de Bruxelles, Blvd. du Triomphe, 1050 Brussels, Belgium}


\begin{abstract}
Recently, it was shown that a theoretical description of nucleation based on fluctuating hydrodynamics and classical density functional theory can be used to determine non-classical nucleation pathways for crystallization (Lutsko, Sci. Adv. 5, eaav7399 (2019)). The advantage of the approach is that it is free of any assumptions regarding order parameters and that requires no input other than molecular interaction potentials. We now extend this fundamental framework so as to describe the dynamics of non-classical nucleation. We use it to study the nucleation of both droplets and crystalline solids from a low-concentration solution of colloidal particles using two different interaction potentials. We find that the nucleation pathways of both droplets and crystals are remarkably similar at the early stages of nucleation until they diverge due to a rapid ordering along the solid pathways in line with the paradigm of "non-classical" crystallization. We compute the unstable modes at the critical clusters  and find that despite the non-classical nature of solid nucleation, the size of the nucleating clusters remains the principle order parameter in all cases supporting a "classical" description of the {\it dynamics} of crystallization. We show that nucleation rates can be extracted from our formalism in a systematic way. Our results suggest that in some cases, despite the non-classical nature of the nucleation pathways, classical nucleation theory can give reasonable results for solids but that there are circumstances where it may fail. This contributes a nuanced perspective to recent experimental and simulation work suggesting that important aspects of crystal nucleation can be described within a classical framework.
 \end{abstract}

\date{\today }
\maketitle

\section{Introduction}
Crystallization  is an every-day
phenomenon for which a fundamental physical description remains elusive. The key first step is nucleation: the formation, via thermal fluctuations, of a sufficiently large cluster of solid as to be thermodynamically stable.  Classical Nucleation Theory (CNT)\cite{Kashchiev},  based on macroscopic concepts such as surface
tension and the assumption that the size of a cluster is the only order parameter, was developed in the first half of the twentieth century continues to define the way we approach the phenomena. However, its extension to crystallization from solution - where both ordering and cluster-building are important - has always been problematic since, after the size of a cluster, any choice of additional order parameters is somewhat arbitrary. Furthermore, modern advances in experimental and simulation capabilities have revealed many so-called non-classical aspects of solid nucleation\cite{Vekilov,DeYoreo, Smeets, CGD_Classical, hColfen,FrenkelSci, Sear, precursors}
 such as multistep nucleation pathways (e.g. crystallization via the formation of dense unstructured
clusters followed by ordering with little mass change) and the frequent
occurrence of nucleation precursors - small clusters that appear to play a role
in nucleation but which are not expected to exist based on the classical considerations. 

Theoretical efforts to go beyond the standard CNT go back a long way and include, e.g., mathematical developments such as
the seminal work of Langer\cite{Langer1,Langer2}, Talkner\cite{Talkner} and others
on multidimensional barrier crossing due to thermal fluctuations (the Kramers
problem\cite{Hanggi}) as well as physically motivated
extensions of CNT to include order parameters beyond the cluster size\cite{Russo, FrenkelSci, Parrinello1, Szabo1, Peters} and attempts to accommodate multistep pathways within its framework\cite{Smeets,CGD_Classical,Gispen}. Recently, it has
been shown that the combination of classical Density Functional Theory (cDFT)
and fluctuating hydrodynamics offers a relatively fundamental, if still
mesoscopic, description of nucleation able to predict a priori the structure
of small clusters (including crystalline clusters)\ and of non-classical
pathways\cite{Lutsko_JCP_2012_1,LutskoHCF}. The advantages of this approach
are that (1) the structure of small clusters is accurately described by cDFT,
even down to molecular length scales\cite{LutskoLam} and (2) no order
parameters are introduced. In particular, there are no a priori assumptions
about crystalline structure, the nature of critical clusters  nor about the form of the nucleation
pathways - all emerge as outputs of a theory based only on a molecular
interaction potential.

  While this approach, which we call Mesoscopic Nucleation Theory,  has many advantages, it remains incomplete as it does not address the aspects of nucleation that
  are of most practical importance and are the focus of most experimental and simulation investigations: namely, the
  dynamics of the process and most particularly, the nucleation rates. The goal of the present work is the fill this gap and to thus
  give a complete theory of nucleation
  that characterizes nucleation pathways, nucleation rates and the emergent
  order parameters resulting in a complete description of nucleation of any system, including crystals, at
  a much more fundamental level than that of CNT. This involves three important contributions. The first is to describe
  the calculation of the unstable modes: the unstable eigenvalue and corresponding eigenvectors of the critical cluster. We note that the
  relevant quantities are not those of the Hessian of the free energy as is often assumed 
  but rather those of the dynamical matrix consisting
  of the product of the kinetic coefficients (determined by the dynamics of fluctuations) and the free energy Hessian. In our examples, we demonstrate the remarkable
  result that, taking this into account,  the natural order parameter - i.e. the dynamics projected onto the unstable mode - corresponds almost entirely to the mass of the cluster for the nucleation
  of both liquid \emph{and crystalline clusters} thus suggesting that both processes are (at least dynamically) ``classical''. This is puzzling as the pathways for the solid clusters are clearly
  non-classical, starting first as droplets and with crystalline ordering only forming late in the nucleation process as previously reported\cite{LutskoHCF}.

  Our second contribution, which allows us to make sense of this apparent paradox, is the introduction of what we term the ``kinetic distance'' as the
  natural reaction coordinate of the process. This is a natural distance measure in density space dictated by, and therefore closely related to, the dynamics of the process so that large kinetic distances imply long times
  and small distances short times. We show that when the pathways are expressed in terms of this quantity the pathways leading to both droplets and solid clusters at
  the same thermodynamic conditions are nearly indistinguishable except that the solids show almost instantaneous jumps corresponding the development of crystalline order at constant mass. 
  Thus, the kinetic distance reveals that the dynamics are largely classical (i.e. as assumed in CNT)
  with very short bursts of ordering characterizing the non-classical parts of the process. In our earlier, non-dynamical, work on the nucleation pathways, the kinetic-distance was not known and we instead used
the Euclidean distance in density space which is impossible to interpret physically\cite{LutskoHCF}. The dynamics of the process are therefore dominated by the long ``classical'' periods of mass accumulation with the short non-classical bursts of ordering being of secondary importance. The process thus possesses both classical and non-classical characteristics simultaneously.

  The third and final contribution  is a means of calculating nucleation rates. We do this with two different methods. In the first, we follow the reasoning introduced by Auer and Frenkel\cite{AuerFrenkel} and give an expression that makes use of the information obtained for the unstable eigenvectors and eigenvalues of the critical cluster. This is expected to be reasonable for droplet nucleation which is already an advance beyond previous work. However, we note that for solids, this expression can be misleading if the critical cluster happens to appear during one of the bursts of ordering and so we also introduce a more heuristic expression involving integration along the entire nucleation pathway and demonstrate that it is consistent with the first method for droplets and gives more physically reasonable results for solids too. An incidental, but satisfying result of our analysis is a theoretical expression for the attachment and detachment frequencies for diffusion-limited nucleation that play a central role in CNT but for which no such theoretical expression has previously been developed.

  These elements -  the unstable modes, the kinetic distance and the expressions for nucleation rates - together with the basic framework of cDFT and fluctuating hydrodynamics\cite{Lutsko_JCP_2012_1} and the determination of the nucleation pathway as the most likely path of the stochastic dynamics\cite{LutskoHCF} give all of the tools necessary for a theoretical description of not only homogeneous nucleation of droplets and crystals as studied in this work, but also of heterogeneous nucleation as well. Being based on hydrodynamics, non-equilibrium effects such as flows and temperature gradients can also be described in a natural manner. Our results for nucleation under diffusion show that it is possible for the nucleation pathway to be highly non-classical while, at the same time, the dynamics of nucleation is largely classical, being dominated by that of the classical order parameter (i.e. cluster size). In this way, we challenge of the dichotomy of classical versus non-classical nucleation that is assumed in much of the literature.

\section{Mesoscopic Nucleation Theory} We focus attention here on a two-species
system consisting of a larger species in a bath of small molecules:\ for
example, macromolecules in solution or colloidal systems. Our considerations
are in particular relevant to the important problem of nucleation from solution,
including precipitation of crystals. Treating the small molecules implicitly and
working in the overdamped limit\cite{Lutsko_JCP_2012_1}, the dynamics of the local number density
$n_{t}\left(  \mathbf{r}\right)  $ of the larger species is modelled with the stochastic differential equation%
\begin{equation}
\frac{\partial}{\partial t}\widehat{n}_{t}\left(  \mathbf{r}\right)
=D\boldsymbol{\nabla}\cdot\widehat{n}_{t}\left(  \mathbf{r}\right)
\boldsymbol{\nabla}\frac{\delta \beta F^{\left(  \text{dyn}\right)  }\left[
\widehat{n}_{t}\right]  }{\delta\widehat{n}_{t}\left(  \mathbf{r}\right)
}+\boldsymbol{\nabla}\cdot\sqrt{2D\widehat{n}_{t}\left(  \mathbf{r}\right)
}\widehat{\boldsymbol{\xi}}_{t}\left(  \mathbf{r}\right)  , \label{DK}%
\end{equation}
where the caret indicates stochastic quantities, $D$ is the tracer diffusion
constant for a single monomer of the large species in the bath, $F^{\left(
\text{dyn}\right)  }\left[  \widehat{n}_{t}\right]  $ is the \textit{dynamical}
free energy functional for the large species and $\widehat{\mathbf{\xi}}%
_{t}\left(  \mathbf{r}\right)  $ is Gaussian-distributed white noise,
representing the effect of the bath. Also, $\beta = 1/k_BT$ where $k_{B}$ is
Boltzmann's constant, $T$ is the temperature. This minimal description can be derived
from the microscopic equations of motion by first using projection operator techniques to
obtain fluctuating hydrodynamics\cite{Grabert, Zubarev} and then adding additional
dissipative and fluctuating terms to take account of the bath followed by the
overdamped limit\cite{Goddard}. The local
density appearing in Eq.(\ref{DK}) is then understood to be a coarse-grained density, e.g. the
instantaneous microscopic density averaged over small volumes. What we have termed the "the dynamical
free energy" is usually the thermodynamic free energy of a system constrained (e.g. by the presence of external fields) to have the specified coarse-grained density, which can also be thought of as a local equilibrium free energy. An
interpretation of this model is that density fluctuations on length scales
smaller than the coarse-graining scale are accounted for in the dynamical free
energy and Eq.(\ref{DK}) describes fluctuations of the density on larger length scales. 

Only formal expressions for the dynamical free energy are known and to proceed one must introduce a model. 
For this, we turn to classical Density Functional Theory (cDFT) where one of the main goals is the development of constrained free energy functionals\cite{Evans1979,LutskoReview}. The sophisticated models developed over several decades are routinely used to determine the free energy, structure and thermodynamic properties of inhomogeneous systems including, e.g., solid clusters\cite{LutskoLam,LutskoSchoonen}. The only inputs required are the interaction potentials of the molecules and no a priori assumptions are made about crystal structures, lattice parameters, etc.  Despite their somewhat different formal origins (cDFT is a strictly equilibrium theory), the cDFT free energy models are the best available examples of constrained free energy functionals capable of describing multiphase systems down to molecular length scales and as such are the only feasible option to use in the present context. This can be viewed as being a type of local equilibrium approximation of the kind commonly resorted to in nonequilibrium statistical mechanics. It has been used in a non-fluctuating context in a derivation of Dynamical Density Functional Theory\cite{EvansArcher} and in the present, fluctuating context, its use in the derivation of fluctuating hydrodynamics based on projection operator techniques has recently been discussed by Dur\'{a}n-Olivencia, et al\cite{Miguel}.

\section{Implementation} Our calculations are based on a discretization of this
model. We label the Cartesian grid points with a super-index
$I\longleftrightarrow\left(  i_{x},i_{y},i_{z}\right)  $ and the values of the
density on the grid as $\widehat{n}_{t}^{I}$. The grid indices take on values
$0\leq i_{a}<N_{a}$ for $a=x,y,z$, the super-index has values $0\leq
I<N_{x}N_{y}N_{z}$ and the grid spacing is taken to be $\Delta$. The operator ${\boldsymbol{\nabla}}\cdot\widehat{n}\left(  \mathbf{r}\right)
\boldsymbol{\nabla}$ becomes a matrix which we call $-g^{IJ}\left( \widehat
{n}\right)  $ that plays the role of a position-dependent diffusivity (or, more abstractly, kinetic prefactor; see Appendix \ref{Discretization} for details). The stochastic differential equation
becomes
\begin{equation}
\label{dyn}\frac{\partial}{\partial t}\widehat{n}_{t}^{I}=-\frac{D}{\sigma^{2}%
}g^{IJ}\left(  \widehat{n}_{t}\right)  \frac{\partial\beta\Omega\left(
\widehat{n}_{t}\right)  }{\partial\widehat{n}_{t}^{J}}+\sqrt{\frac{2D}{\sigma^{2}}%
}\mathbf{q}^{IK}\left(  \widehat{n}_{t}\right)  \cdot\widehat{\boldsymbol{\xi
}}_{t}^{K},
\end{equation}
where we use the Einstein convention of summing over repeated indices and we
let $\widehat{n}_{t}$, with no indices, represent the entire collection of
values $\widehat{n}_{t}^{I}$. The first term on the right now refers to
$\Omega\equiv F - \mu N$ which is the result of our use of fixed-density
boundary conditions to model an open system (see Appendix \ref{BC} for more details). In the
noise term, the amplitude and noise are both 3-vectors at each lattice point.
The amplitude of the noise is related to the coefficient of the gradient via
$\mathbf{q}^{IK}\cdot\mathbf{q}^{JK}=g^{IJ}$, which is a
fluctuation-dissipation relation assuring that in equilibrium, the system
obeys a canonical distribution $P\left(  n\right)  \sim\exp\left(
-\beta\Omega\left(  n\right)  \right)  $. Finally, and importantly, we will
consider here only open systems for which the density on the boundaries of the
computational cell is held constant:\ this is equivalent to setting the
chemical potential for the system. In this representation a pure solution
corresponds to constant density throughout the system\cite{Lutsko_JCP_2012_1}.

As is commonly done, we will  restrict ourselves to the weak-noise limit
which in some sense corresponds to low temperatures\cite{Lutsko_Duran_2015}
for which the following properties have been
established\cite{Lutsko_JCP_2012_1}:\ if the system makes a transition from
one local minimum of the free energy to another, then the most likely path
(MLP) (given the stochastic dynamics)\ passes through a saddle point and
corresponds to the two steepest descent paths that result from perturbing the
system at the saddle point in the direction of the unstable eigenvector (in
positive and negative directions). In this way, the concept of the critical
cluster (a saddle point on the free energy surface) and its critical role in
the phase transition emerges naturally. The critical cluster
$n^{\ast}$ is  defined by
\begin{equation}
\left.  \frac{\partial\beta\Omega\left( n\right)  }{\partial n^{J}}\right\vert
_{n^{\ast}}=0,
\end{equation}
and expanding the stochastic equation about this value gives%
\begin{equation}
\frac{\partial}{\partial t}\delta\widehat{n}_{t}^{I}=-\frac{D}{\sigma^{2}}%
g^{IJ}\left( n^{\ast}\right)  \left(  \frac{\partial^{2}\beta\Omega\left(
n\right)  }{\partial n^{J}\partial n^{K}}\right)  _{n^{\ast}}\delta\widehat
{n}_{t}^{K}+\sqrt{\frac{2D}{\sigma^{2}}}\mathbf{q}^{I J}\left( n^{\ast}\right)
\cdot \widehat{\boldsymbol{\xi}}_{t}^{J},
\end{equation}
so that the unstable direction is a solution of the eigenvalue problem
\begin{equation}
g^{IJ}\left(  n^{\ast}\right)  \left(  \frac{\partial^{2}\beta\Omega\left(
n\right)  }{\partial n^{J}\partial n^{K}}\right)  _{n^{\ast}}v^{K}=\lambda
v^{I}%
\end{equation}
for $\lambda<0$. The matrix in parenthesis is the Hessian of the free energy
and the product of the kinetic factor times this matrix will be referred to
henceforth as the dynamical matrix. Because the dynamical matrix is real, but not
symmetric, there is, for each eigenvalue, a corresponding left eigenvector
that solves%
\begin{equation}
u_{I}g^{IJ}\left(  n^{\ast}\right)  \left(  \frac{\partial^{2}\beta
\Omega\left(  n\right)  }{\partial n^{J}\partial n^{K}}\right)  _{n_{t}^{\ast
}}=\lambda u_{K}%
\end{equation}
and the collection of left and right eigenvectors form a bi-orthogonal set.
While distinct, it is easy to see that the two eigenvectors are related by
$u_{I}g^{IJ}\left(  n^{\ast}\right)  \propto v^{J}$. In fact, as discussed in Appendix \ref{Eigen}, we will use a normalization throughout this work according to which
$u_{I} g^{IJ} = v^{J}$ and $u_{I} v^{I} = 1$. It has been discussed
elsewhere\cite{Lutsko_JCP_2012_1} that the structure of the stochastic
equation imposes a Riemannian geometry on density space, with the matrix
inverse of the diffusivity matrix, denoted with subscripts as $g_{IJ}$,
playing the role of the metric on the space. For this reason, we use the
standard notation of covariant and contravariant vectors related via the
metric. This just means that for any (covariant) vector with lower indices,
$v_{I}$ there is a corresponding (contravariant) vector with upper indices
defined as $v^{I} = g^{IJ} v_{J}$.

\section{Nucleation of droplets and solids: order parameters and pathways.} 
We will first illustrate this framework with calculations for the nucleation of dense
droplets from a low-concentration, or ``weak'',  solution of particles interacting via a Lennard-Jones (LJ)
potential and via a more short-ranged pair interaction introduced by Wang et al  to model
colloidal interactions\cite{Wang} and which we refer to as the WHDF potential (the explicit forms of both potentials are given in Appendix \ref{Details}). The WHDF interaction gives generally lower surface tensions than the LJ thus allowing us to sample both the high (with LJ) and low (with WHDF) supersaturation regimes at similar computational expense. As stated above, what we have called the dynamical free energy is modeled by a cDFT functional, specifically 
the explicitly stable Fundamental Measure Theory model\cite{esFMT} with
the attractive part of the potential described by a mean field term. Our grid
has 64 points in each dimension for the LJ potential and 128 for the WHDF potential giving more than 250 thousand grid points overall in the former case and more than 2 million in the latter. The discretization and constant-density boundary conditions used to
mimic an open system are described in Appendices \ref{Discretization} and \ref{Eigen}. While the
dynamical matrix is far too large to be calculated explicitly, it turns out that the
structure of the cDFT model allows for the computation of its product with any
given vector to be performed efficiently using fast Fourier Transforms. This
is enough to allow the use of sophisticated libraries for the calculation of a
limited number of eigenpairs (including both left and right eigenvectors), and
our calculations use SLEPC\cite{SLEPC1,SLEPC2,SLEPC3,SLEPC4}.

\begin{figure*}
\includegraphics[width=0.45\linewidth]{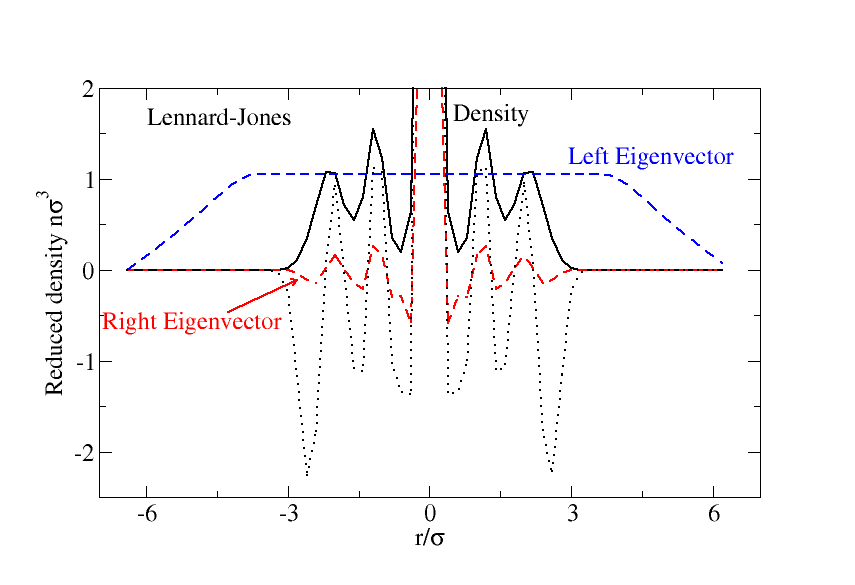}
\includegraphics[width=0.45\linewidth]{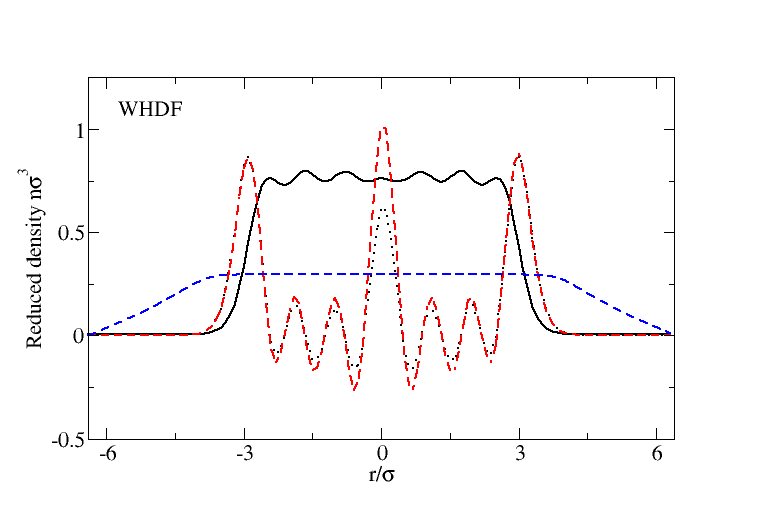}
\caption{The black line shows that density in a critical droplet as a function of distance from the center, for an open Lennard-Jones fluid with temperature $k_BT/\epsilon = 0.3$ and
supersaturation $S=4.00$ (left panel) and for a WHDF fluid with $k_BT/\epsilon = 0.275$ and $S=0.51$ (right panel), where $\epsilon$ and $\sigma$ are the energy and length scales of the respective potentials. The oscillations are signs of packing of the fluid due to the confinement and the large peak in the Lennard-Jones density indicates a high probability that a particle is in the center of the droplet. The dashed lines show the left and right eigenvectors for the unstable mode of the dynamical matrix and the dotted line shows the same for the Hessian of the free energy, with no
dynamics (left and right eigenvectors are therefore the same). The right eigenvectors (and those based on the Hessian) reflect the structure of the droplet whereas the (dynamic) left eigenvector is very simple, corresponding to a probe of the
mass of the droplet, a result only possible because of the mass-conserving dynamics. Note that the droplets are spherically symmetric, so the Figures give complete information about the various quantities. }
\label{eigvec}
\end{figure*}

Critical
clusters were constructed (as described in Appendix \ref{Details}) for the Lennard-Jones system under conditions of high supersaturation ($S\equiv \beta \Delta \mu = 4.00$ where $\Delta \mu$ is the excess chemical potential) and conditions of low supersaturation ($S = 0.51$) for the WHDF system. Then, the largest several (up to 10)
eigenvalues and eigenvectors were calculated. In the examples presented here,
as well as many others, we find a single unstable (negative) eigenvalue.
Figure \ref{eigvec} shows the droplet critical clusters as well as the left and
right eigenvectors in the unstable direction. The density of the critical
clusters  oscillates which is typical any time a liquid with strong
short-ranged repulsion is confined:\ it represents layering of the fluid in
alternating shells of high and low density. In the LJ droplet, a large density at the center of the cluster simply indicates a high probability of finding a particle there. The right eigenvectors, the
unstable directions in density space, are complicated, with a lot of structure
corresponding to that of the density itself. However, the left eigenvectors are
remarkably simple, being nearly flat within the clusters and then
decreasing outside the clusters until vanishing at the boundary. (That it is not completely devoid of structure is shown in the Supplementary Information\cite{supp}.)

The meaning of this simplicity becomes clear when one considers two points. First, since the
eigenvectors of the dynamical matrix form a complete set, any local density,
$n$, can be written as an expansion of the form $n=\sum_{\alpha}c^{\left(
\alpha\right)  }v^{\left(  \alpha\right)  }$ where the index $\alpha$ labels
the different eigenvectors (so it is not lattice index) and the coefficients are
calculated from the corresponding left eigenvectors, $c^{\left(
\alpha\right)  }=v_{I}^{\left(  \alpha\right)  }n^{I}$. We let $\alpha=-$
correspond to the unstable (negative eigenvalue) mode, so that the projection
of the density onto the unstable direction is the coefficient
$c^{\left(  -\right)  }$. The second observation is that the total mass of the
system is calculated as the sum of the density at each point multiplied by the
volume element, $\sum_{I}n^{I}\Delta^{3}$, and this can be written as $M_{I}n^{I}$
where the covariant vector $M_{I}$ has all elements equal to $\Delta^{3}$ which
would simply be a horizontal line in traversing the system in Fig.
\ref{eigvec}. The excess mass of the cluster is then $\Delta N=$ $M_{I}\left(
n^{I}-n_{0}^{I}\right) $ where $n_{0}^{I}$ is the uniform density of the mother phase. Since the left eigenvector is (nearly) constant within the cluster,and
the density outside the cluster is virtually the same as in the mother phase,
it follows that $v_{I}^{\left(  -\right)  }\left( n^{I}-n_{0}^{I}\right) $ is very nearly proportional to the excess mass  as well. Thus, the projection of the local density
onto the unstable direction is, up to an irrelevant additive constant, a
measure of the excess mass of the cluster. Since the unstable direction is
also the MLP for nucleation, the identification of the projection of the
density onto this direction as the order parameter is intuitively appealing.
Furthermore, Berezhkovskii and Szabo\cite{Szabo1} (see also the discussion in
ch. 18 of Peters\cite{Peters}), have shown that this is the only choice of
order parameter that leads to the correct nucleation rate in a one-dimensional
description of nucleation. All of this together leads to the conclusion that
the emergent order parameter at the critical cluster is, in both cases, the excess mass of the cluster as assumed in CNT. This conclusion is reassuring since one expects CNT to describe droplet nucleation. 

\begin{figure*}
\includegraphics[width=0.3\linewidth]{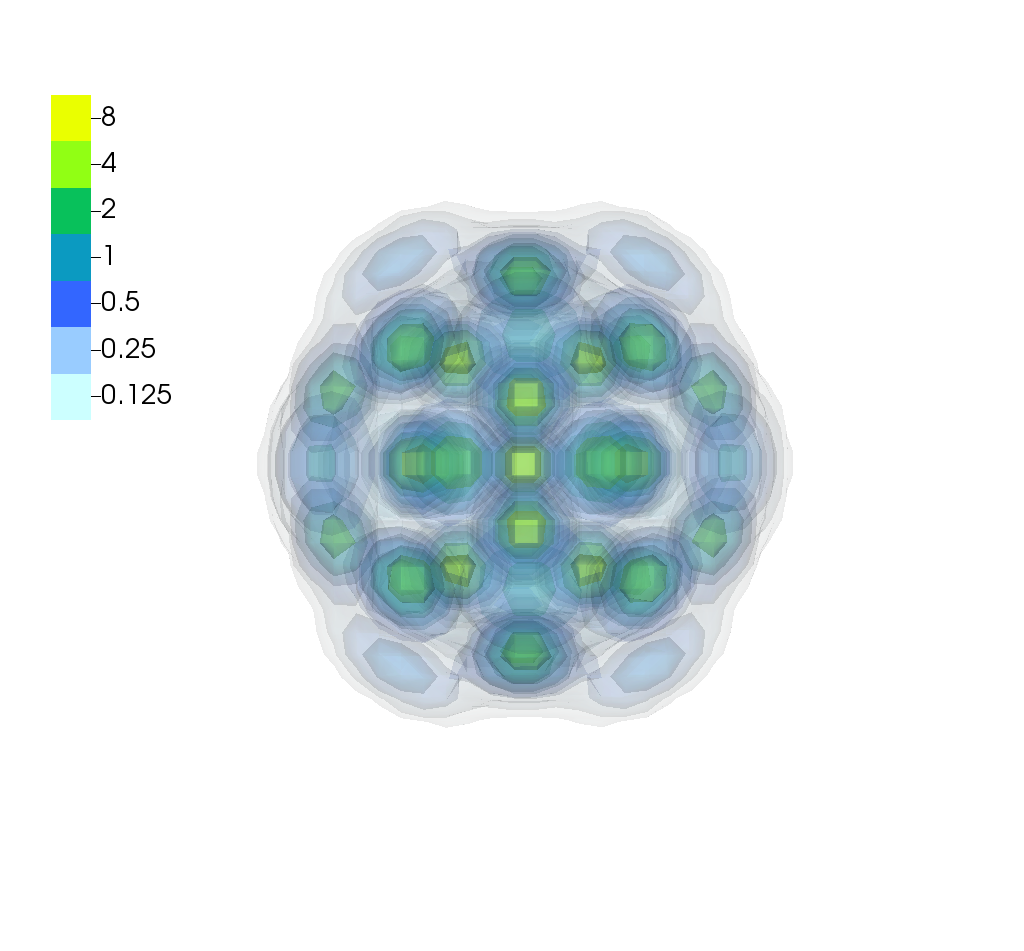}
\includegraphics[width=0.3\linewidth]{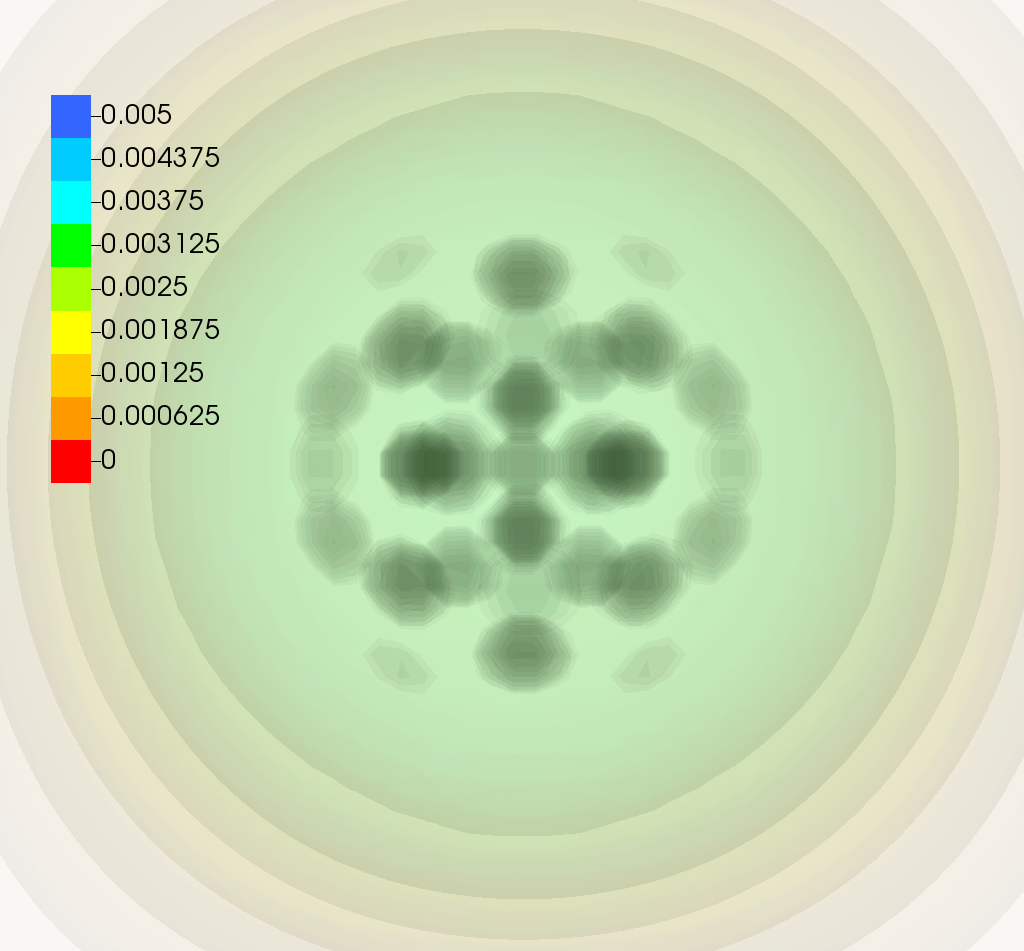}
\includegraphics[width=0.3\linewidth]{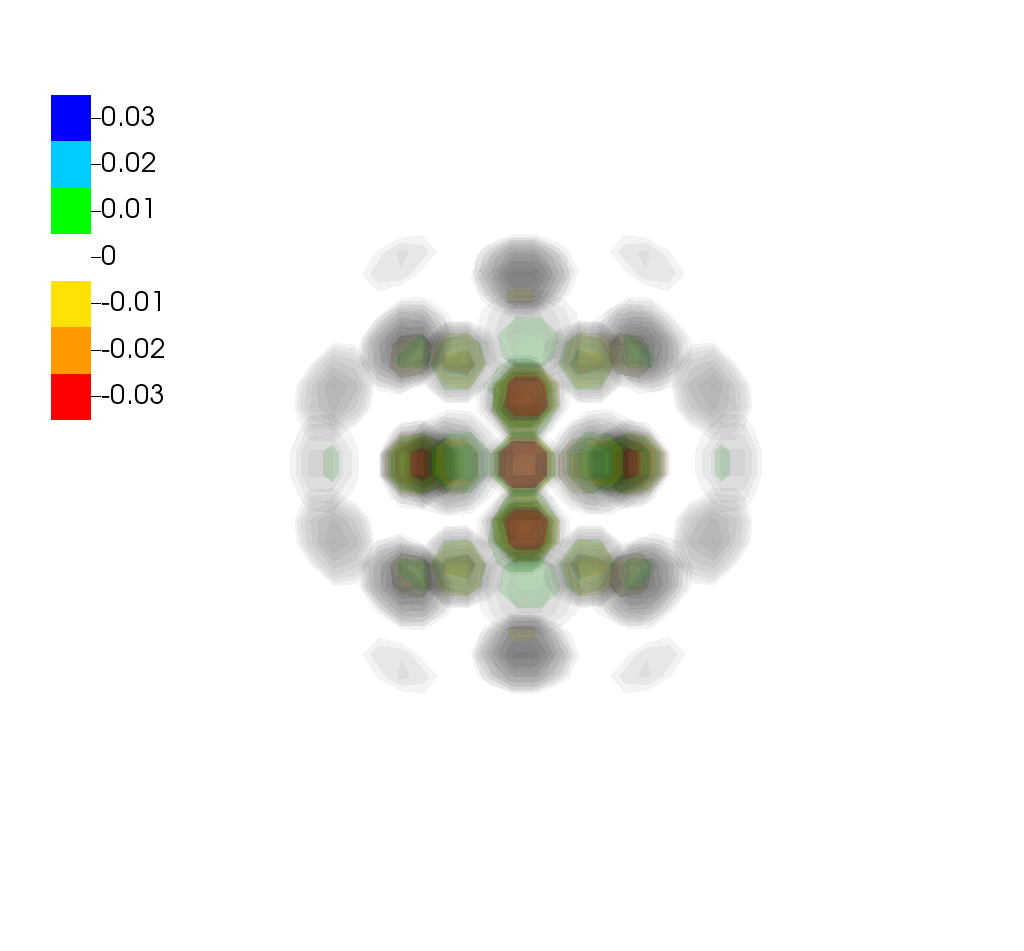}
\caption{ Local density of the solid-like critical cluster (left, displayed on a log-scale) and the left and right eigenvectors (center and right, both linear scales) for the unstable mode of the LJ cluster. The cluster has an excess mass of about 40 particles of which about half are localized into fixed ``atoms'' and about half in a fluid state. In the eigenvector images, the local density is represented in grey
and the eigenvectors in color. The eigenvectors are qualitatively very similar as for the droplet: the left eigenvector is nearly constant inside the solid cluster while the right
eigenvector is similar to the density as evidenced by the fact that one only observes color on the localized grey sites.}
\label{solid_eigvecs}
\end{figure*}

Again using the procedures described in Appendix \ref{Details}, we have  generated solid clusters under the same thermodynamic conditions as for the droplets. As previously observed\cite{LutskoHCF}, crystallization pathways involve an initial critical cluster followed by a sequence of shallow minima and maxima corresponding to the formation of complete shells of solid as the cluster grows.
Figure \ref{solid_eigvecs}
shows the \emph{first} critical cluster and the unstable eigenvectors associated with it for the LJ potential (see the Supplementary Information\cite{supp} for corresponding images for the WHDF case). 
Roughly, half the mass is localized into ``particles''
and half is delocalized liquid-like fluid.  We recall that in this formalism,
there is no explicit representation of colloidal particles or of crystal
lattices and that all such structure observed forms spontaneously during
minimization of the free energy so that ``particles'' are really just
locations where the local density (the probability of finding a colloidal
particle) is very high, surrounded by a region with very low probability thus
indicating a localized particle. In this cluster, as well as in the WHDF case, these "particles"
are located at the vertices of an icosahedron, with additional peaks in front
of each one of its 20 facets. This hcp-like stacking of particles on the
icosahedron facets is the first shell of what is sometimes referred to as an
``anti-Mackay'' structure. Structures with icosahedral symmetry have been
observed in super-cooled liquids and glasses\cite{Andersen,Hirata,Singh} and
can be obtained when a liquid crystallizes in a confined
environment\cite{de-Nijs,Mbah}. The eigenvectors are shown in Fig.
\ref{solid_eigvecs} as 3-dimensional contour plots overlaid on the cluster
image. One clearly sees the same structure as for the droplet: the right
eigenvectors reflect the density of the cluster whereas the left eigenvector
is, quite surprisingly given the complexity of structure, almost constant within the cluster,  as
in the case of the droplet. Once again, this suggests that the order parameter is closely related to the excess mass of the clusters.

\begin{figure*}
\includegraphics[width=0.5\linewidth]{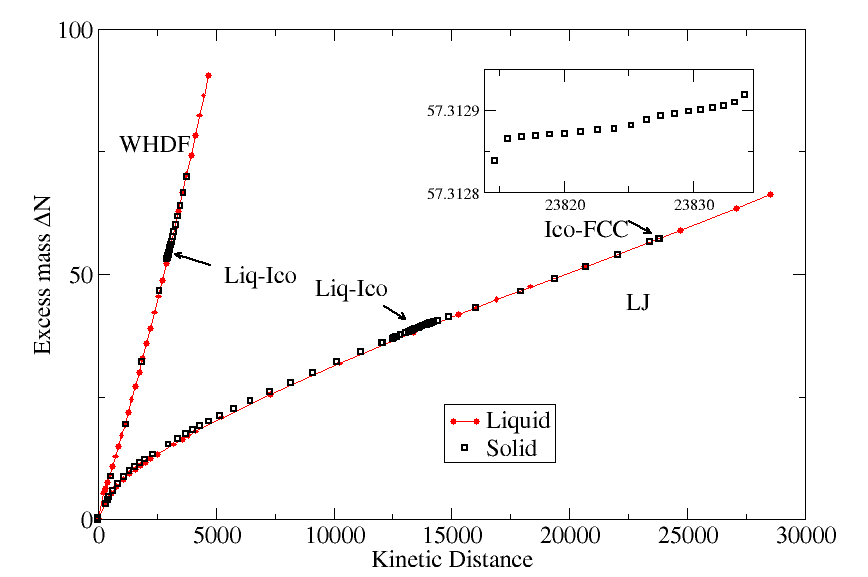}%
\includegraphics[width=0.5\linewidth]{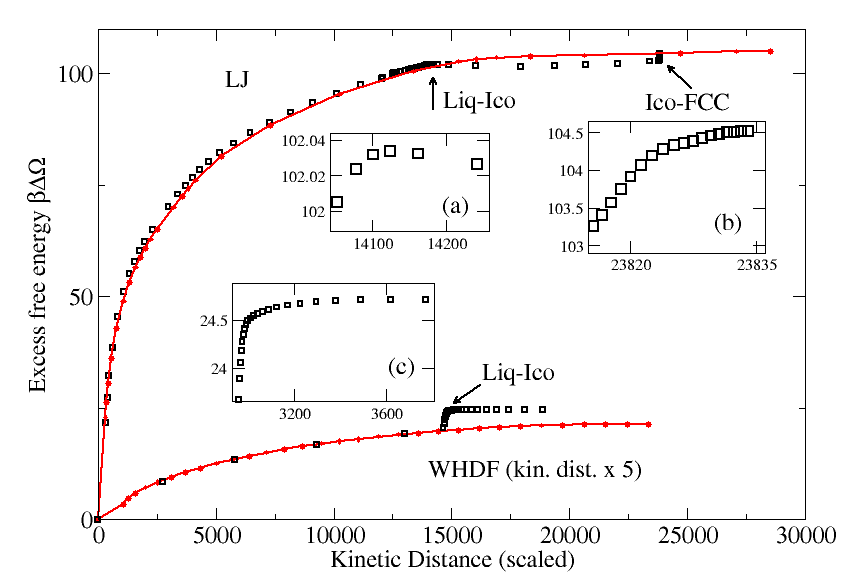}
\includegraphics[width=0.5\linewidth]{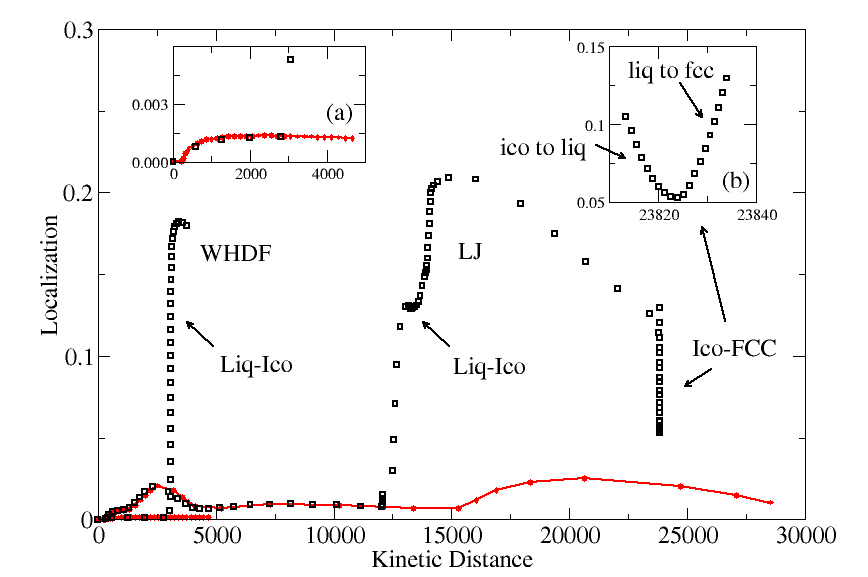}
\caption{ Excess mass (total number of particles compared to the initial uniform state, left panel), 
 excess free energy (center panel) and "localization" (right panel) as functions of kinetic distance along the most likely paths for both the solid cluster and the liquid droplet for both potentials. Black symbols are results for the "solid" pathways and red lines are the droplet pathway. The kinetic distance for the WHDF system is multiplied by a factor of 5 in the middle panel so as to improve visibility. All pathways end in a critical cluster; in both solid pathways, an initially liquid-like droplet first transforms into an icosahedral cluster. In the case of LJ, this is followed by a solid-solid transition to the FCC structure. The localization is a measure of order within the cluster (as described in Appendix \ref{order})  and shows that the near discontinuity in the free energy corresponds to a sharp increase into solid-like localization of the density and then the reorganization into an FCC structure. The left and center panels both include insets showing a detailed view of the reorganization of the liquid droplet into an icosahedral solid (inset a). The center panel also shows similar zooms of the ico-fcc transition in LJ (b)  and the liquid-ico transition in the WHDF system (c). The insets in the right hand panel show (a) a close-up of the divergence between the droplet and solid pathways for the WHDF interaction and (b) the extreme point of the transition from icosahedral order to FCC order via a dis-ordered, fluid-like intermediate state.}
\label{pathway}
\end{figure*}

We have calculated nucleation pathways (e.g. the most likely path) from the initial low-concentration state to the various critical clusters using the string method\cite{LutskoHCF}. In brief, one creates an initial guess at
the path consisting of some number (we typically use 30) images of the system with the
density interpolated between the initial, uniform, state and the final
critical cluster which is then relaxed based on the deterministic dynamics,
under the constraint that the distance between the images remains fixed. 

To present the results, we introduce  a natural measure of distance in density space. Note that the deterministic dynamics is driven by the
gradient of the free energy, combined with the kinetic factor, 
$g^{IJ}(n)$ in Eq.(\ref{dyn}). This can be interpreted as gradient descent in a
curved space in which one measures the distance between a density distribution
$n^{I}$ and another $n^{I}+dn^{I}$, infinitesimally close, as $ds^{2}= dn
^{I}g_{IJ}dn^{J}$ (we will call this dimensionless quantity the ``kinetic distance'' and details concerning its calculation are given in Appendix \ref{string_details}). 

The excess mass (number of particles) and free energy of
the clusters as they evolve along the nucleation pathway are shown in Fig.
\ref{pathway} using the kinetic distance along the pathway as the independent variable. The nucleation pathway for the LJ cluster is complex: first an icosahedral critical cluster is formed followed by a solid-solid transition to an FCC crystalline lattice. For the WHDF interaction, our calculation terminates at an icosahedral structure and further calculations
 would be necessary to determine if an eventual transition to a close-packed structure occurs (as one would expect).
 
 These pathways exhibit several remarkable features. In particular, the 
kinetic distance along the pathways corresponds monotonically, and almost linearly, to excess mass in
both cases. Even more surprising, the excess energy of the eventual solid and liquid clusters are almost
identical until, at a given point the (eventual) solid-like cluster shows an
abrupt jump in free energy, almost a discontinuity. During this processes, the
excess mass is almost constant and the main difference is that the
localization into ``particles'' occurs. This leads to the unexpected
conclusion that while ordering does occur independently of mass accumulation,
in a kind of three-step process of mass increase, then ordering, then mass
increase, the ordering is very fast (in the sense that the kinetic distance between
the ordered and unordered clusters is very small). Nevertheless, the unstable
eigenvalues at the critical cluster differ significantly for the two
processes as can be seen from the values given in Table \ref{tab:eigenvalues} . The table also gives the unstable eigenvalues for the corresponding free energy Hessians and illustrates the differences between the two calculations. 
\begin{table*}
    \centering
    \begin{tabular}{c|ccc|ccc}
         & \multicolumn{3}{c}{$\boldsymbol\lambda^{(-)}\boldsymbol\sigma^{-6}$ {\bf  (Free Energy Hessian)}}&\multicolumn{3}{c}{$\boldsymbol\lambda^{(-)}$ {\bf (Dynamical Matrix)}} \\
         & Droplet & Solid & Ratio &  Droplet &  Solid &  Ratio  \\\hline \hline
         LJ  (Ico)        &     $-0.00505$&  $-0.00280$& $1.8$ &    $-1.61\times10^{-5}$&  $-6.88 \times10^{-3}$& 0.0024\\
         LJ  (FCC)        &     $-0.00505$&  $-3.71 \times 10^{-5}$& $136$&    $-1.61\times10^{-5}$&  $-5.51 \times10^{-2}$& $0.00029$\\
         WHDF&  $-1.93 \times10^{-4}$&  $-6.44 \times10^{-6}$&  $30$&  $-9.42 \times10^{-4}$&  $-1.75 \times10^{-4}$&  $5.4$\\
    \end{tabular}
    \caption{Unstable eigenvalues of the Hessian of the free energy and of the dynamical matrix for the two interaction potentials. For the dynamical matrix, the eigenvalues are dimensionless while for the Hessian they have units of $(length)^6$ and are expressed in terms of the length scale of the potential. The ratio of the droplet to the solid are also given in each case and illustrates how different are the static and dynamic values. }
    \label{tab:eigenvalues}
\end{table*}

\section{Nucleation rates} Classical Nucleation Theory is based on the
assumption that the number of particles in a cluster is a good order parameter
for the transition from vapor to liquid or to solid. One expects that in the
general context of MeNT, the order parameter should in some sense correspond
to movement along the nucleation pathway which would mean that at the critical
cluster, it would involve the projection of the density onto the unstable
direction, $v_{I}^{(-)}n^{I}$ and the fact that $v_{I}^{(-)}\left(
n^{I}-n_{0}^{I}\right)  \propto\Delta N$ indicates that the excess mass is
indeed a good order parameter. The same assumption underlies the
\textquotedblleft parameter-free\textquotedblright\ estimations of nucleation
rates from simulation developed by Auer and Frenkel\cite{AuerFrenkel}. The
idea is to begin with the CNT expression for the nucleation rate,
\begin{equation} \label{JCNT}
J=n_{0}f(N^{\ast})\sqrt{\frac{1}{2\pi}\left|\beta\Omega^{\prime\prime}\left(
N^{\ast}\right)\right|  }\;\;e^{-\beta\Delta\Omega^{\ast}}
\end{equation}
where (in our version) $\Omega\left(  N\right)  $ is the grand-canonical free
energy (which in CNT depends only on the order parameter, namely the mass of
the cluster), $\Delta\Omega^{\ast}=\Omega\left(  N^{\ast}\right)  -\Omega_{0}$
is the excess free energy of the critical cluster with $\Omega_{0}$ that of
the mother phase, $f\left(  N^{\ast}\right)  $ is the rate of attachment of
monomers to the critical cluster (which, by definition of the critical
cluster, is exactly equal to the rate of detachment of monomers) and $n_{0}$
is the density (concentration)\ of the mother phase. Surprisingly, the
CNT attachment rate for diffusive systems is not agreed upon: although some works
use the Smolochowski rate for coalescence\cite{Kashchiev}, this is not
well-accepted\cite{Peters} and more often one resorts to ill-defined concepts
such as typical jump distance and typical jump time to make estimates. Auer
and Frenkel evaluated these various elements directly from simulation and
then used the values to get a model-free determination of the nucleation rate
(assuming that the size of the cluster is the correct order parameter). 

We show in Appendix \ref{Rates} that the attachment and detachment frequencies at the critical cluster
can be extracted from our model with the result
\begin{equation}
f(N^{\ast})=g(N^{\ast})=\frac{D}{\sigma^{2}}\left(  M_{J}v^{\left( -\right)
J}\right)  ^{2},
\end{equation}
and that this results in the CNT nucleation rate 
\begin{equation}
J=n_{0}\frac{D}{\sigma^{2}}\left\vert M_{J}v^{\left(  -\right)  J}\right\vert
\sqrt{\frac{\left|\lambda^{\left(  -\right)  }\right|}{2\pi}}e^{  -\beta\Delta
\Omega^{\ast}}  \label{rate}%
\end{equation}
which involves only quantities accessible from the  model free energy functional and the corresponding dynamical matrix. Evaluations using this are given in Table \ref{tab:rates}. 
In the case of the WHDF interaction, the kinetic prefactor for the solid and the droplet are almost the same and the difference in the rates of the processes is entirely due to the Arrhenius factors. In contrast, the kinetic factors for the LJ solids are much larger (indicating faster processes) than for the droplet, which seems paradoxical since the solid pathway involves the same accumulation of mass as for the droplet plus the additional time needed for the ordering: even if the latter occurs very fast, the mass accumulation should be the same. 

To further understand these results, we have also
developed a more elaborate evaluation of the rates based on a one-dimensional theory derived from our analysis and that involves integration along the
entire nucleation pathway (see Appendix \ref{Heuristic}, Eq.(\ref{J1})). Results using this expression are also given in the table. We cannot claim that this full-path value is more accurate than the CNT result given the limited resolution of the nucleation pathway using the string method. Nevertheless, it is interesting to note that for WHDF, the full-path rate is very similar to the CNT result, differing only by a factor of two. However, while the kinetic factor for the LJ droplet is again similar to the CNT value, the values for the solids are quite different from their CNT values and indeed of similar magnitude as for the droplet. Referring to Fig. \ref{pathway} and Table \ref{tab:eigenvalues} provide some insight. In the case of the WHDF solid, the critical cluster appears after a period of mass accumulation that in turns occurs after the abrupt ordering: the eigenvalues being computed and used in the CNT rate expression are therefore characteristic of the mass-accumulation process. In contrast, the LJ solid critical clusters both appear during the ordering process so that the eigenvalues being computed are related to the time-scale of the ordering: the large dynamic eigenvalues obtained for these cases indicate the relative speed of the ordering. Only by integrating over the entire path is the much slower mass-accumulation accounted for, in which case the kinetic factors are much more comparable to that of the droplet.   

Finally, we note that the fact that the \emph{for identical thermodynamics conditions} the nucleation rate for droplets in the WHDF system is much higher than for direct nucleation of crystals suggests the preferential formation of the latter with the possibility that crystal nucleation may then occur within the droplets. In contrast, no such preference is suggested for the LJ system thus implying the preferential formation of crystals directly from the solution.

\begin{table*}
    \centering
    \begin{tabular}{c|cc|ccc|ccc}
         & \multicolumn{2}{c}{$\boldsymbol \beta \Delta \Omega^\ast$} & \multicolumn{3}{c}{{\bf CNT Rates }}&\multicolumn{3}{c}{ {\bf  Full Path Rates }} \\
         & Droplet & Solid & Z-Droplet& Z-Solid& $\frac{J_{\text{droplet}}}{J_{\text{Solid}}}$&  Z-Droplet &  Z-Solid &  $\frac{J_{\text{droplet}}}{J_{\text{Solid}}}$\\\hline \hline
         LJ (Ico)         & 104.9&102.0 &     $1.76\times10^{-9}$&  $1.65 \times 10^{-6}$& $6 \times 10^{-5}$&    $3.38 \times 10^{-10}$&  $5.56 \times 10^{-10}$& $0.03$\\
          LJ (FCC)        & 104.9&104.5 &     $1.76\times10^{-9}$&  $3.03 \times 10^{-7}$& $4 \times 10^{-3}$&    $3.38 \times 10^{-10}$&  $1.74 \times10^{-9}$& $0.13$ \\
         WHDF& 21.4 & 24.7 &  $9.24 \times10^{-5}$&  $1.67\times10^{-4}$&  $15$&  $ 5.33 \times10^{-5}$&  $4.64 \times10^{-5}$&  $31$\\
    \end{tabular}
    \caption{Calculated nucleation rates for the LJ icosahedral saddle (Ico), the LJ FCC critical cluster and the WHDF critical cluster. The table also shows the free energy barriers, the prefactors $Z \equiv J/e^{-\beta \Delta \Omega}$ of the exponentials in Eq.(\ref{rate}) and the ratio of the full rates calculated from the same expression. The final section of the table shows the Z-factors and the ratios of the full rates as calculated from the heuristic one-dimensional model involving the full nucleation pathway. }
    \label{tab:rates}
\end{table*}

\section{Discussion} We have shown how the powerful combination of cDFT and
fluctuating hydrodynamics can give unique insight into the process of
crystallization at the microscopic level. Our results highlight the absolute necessity of a realistic dynamical
description of fluctuations. In the present case, if one simply computed the
unstable mode of the free energy, it would lead to the erroneous conclusion
that the solid and droplet clusters are governed by very different order
parameters, reflecting their very different density distributions, and that
the order parameters having little physical interpretation. In contrast, the
fact that MeNT is based on hydrodynamics which conserves mass leads both to
different left and right eigenvectors and to the critical and surprising fact
that the projection of the density onto the unstable mode is simply a measure
of the excess mass of the system.

For the examples of a solid-like critical clusters discussed here, it was found
that the development of order, which is a necessary part of the formation
of crystalline structure seems to have little effect on the kinetics of the
process and that, contrary to most expectations, taking account of a separate
order parameter corresponding to crystalline order is unnecessary. This
provides independent support for the idea of using CNT-like expressions for
the nucleation rate for crystallization, as underlies the original work of
Auer and Frenkel\cite{AuerFrenkel}, and the subsequent works based on it. However,  we found that if the critical cluster appears during the ordering process, the kinetics sampled in its neighborhood are not typical of mass accumulation and can therefore give deceptively large kinetic factors.  In fact, our results suggest that the kinetic factors for the solid should in general be comparable to that for a droplet. 

There are long-standing and important descrepencies between nucleation rates as determined from experiment, simulation and theory and the subject is quite complicated as the details of each comparison are specific to the particular techniques being discussed. Well-known examples are the comparison between experiment and simulation for hard-spheres which is reviewed with a proposed resolution by W\"{o}hler and Schilling\cite{Schilling} and that in protein crystallization between experiment and theory\cite{Peter} which drove a lot of the interest in non-classical pathways in those systems. Our work concerns precipitation from solution and so is most similar to the protein system, with the hard-spheres being an example of crystallization from the melt. In any comparison to theory, what is almost always meant is a comparison to CNT but then one faces the question as to where the necessary input comes from. Experiment can give values for the barriers but as stated in Ref.  \onlinecite{Peter2} (where the kinetic prefactor is called $A$), ``There have been attempts to theoretically derive an expression for this coefficient for nucleation from solution. In all cases, the final formulas for A contain variables that are often impossible to determine independently.'' and so recourse is often made to fitting the CNT expressions to extract the coefficient. Thus, only the prediction of the barrier, and the corresponding Arrhenius factor, can really be tested. Our work gives, first, the means to fully evaluate the kinetic prefactor (via the attachement frequency and the unstable eigenvalue) and, second, by its use of cDFT, a much more fundamental determination of the free energy barrier than that of CNT, indeed one which has already been shown to agree well with simulation\cite{LutskoIII}. It thus offers a possible pathway to meaningful and completely independent comparisons of theory and experiment.

{More broadly, while the aim of the present work has not been to match up to any particular experiment or simulation, our results raise some general points that could nevertheless be relevant to such comparisons. In our examples, we have found that there are two different dynamical regimes active during nucleation: mass accumulation, which is diffusive and relatively slow, and ordering, which seems to occur on much shorter time scales. Any method, e.g. simulation-based seeding methods that use the techniques introduced by Auer and Frenkel\cite{AuerFrenkel}, determine the rates based on the dynamics in the close neighborhood of the critical cluster. If this occurs during the ordering, as happened in one of our examples, one will sample the fast ordering dynamics which would not accurately reflect the rate of the processes, which, in our view, is dominated by the much slower mass accumulation phases.

Despite our contention that the dynamics appear largely classical in that they are dominated by the dynamics of the CNT order parameter, we emphasize that the nucleation pathways for crystallization
studied here are multistep in nature: formation and growth of a liquid-like
droplet - which seems indistinguishable whether the pathway results in a droplet, a crystalline cluster  or an amorphous cluster -  followed by rapid localization into solid-like regions, followed
once more by mass accumulation until criticality is reached. This is
consistent with the pathways described previously for Lennard-Jones
systems\cite{LutskoHCF}. The suddenness of the ordering was not evident in the
earlier work because paths were plotted as functions of the \emph{Euclidean}
distance in density space and not the \emph{kinetic} distance that is relevant for
the dynamics (see additional figures in Supplementary Information\cite{supp}). Our unexpected conclusion is that despite this multistep
process, the nucleation rate appears to be quite insensitive to the ordering -
other than via the jump in free energy - and so can be described in the
language of CNT. This provides a line of evidence, completely independent of experiment and simulation, for the idea\cite{Smeets,CGD_Classical,Gispen}  that while the details of the nucleation pathways of crystallization may be (or, perhaps, usually are) highly nonclassical - with independent phases of mass accumulation and ordering and possibly multiple intermediate structures - the dynamics is dominated by the diffusive, and therefore slow, rate of mass accumulation which is occasionally interrupted by brief periods of ordering or re-ordering thus giving it a mostly classical appearance. Thus,  we challenge the dichotomy of classical versus non-classical processes and show that different aspects of nucleation can be one or the other at the same time. In this sense, a fully classical or fully non-classical nucleation process may be the exception.

We note also the important role played by the icosahedral clusters which seem to occur frequently as the first manifestation of incipient solid-like structure. Strikingly, in other calculations that we will describe elsewhere and in which the droplet has higher free energy barrier than the solid  - we see a droplet go through an icosahedral intermediate step which then transforms back to a droplet. Our work seems to indicate in general very complex landscapes with many local maxima and minima and correspondingly complicated pathways. 

Any theoretical approach to this
phenomenon is going to involve heuristic elements due to the underlying
complexity and intrinsically multiscale nature of the problem and ours is no
exception. The principle assumption is our use of cDFT free energy functionals as the dynamic free energy. This is an assumption that could be explored at a formal level but here we only note that in practice, the most common cDFT free energy functionals are just mean field models and as such, it could be argued that it makes more sense to use them as we do here, as part of a dynamic description, than as true thermodynamic free energies, as in cDFT, which should contain renormalization effects that go beyond the mean field. 

The applications of our framework are much broader than the specific results
given here. In principle, there is no necessity to work in the over-damped
limit and, by keeping the full hydrodynamic description, the effect of flows,
heat transport and other important physical phenomena could be studied as
well. Exactly the same concepts can also be used in the study of heterogeneous nucleation which is of more practical importance than homogeneous nucleation but also more problem-specific. It is also possible to determine the MLP without resorting to the weak
noise limit\cite{Lutsko_JCP_2012_1} which would allow for nucleation without necessarily 
passing through the critical cluster\cite{Lutsko_Duran_2015}. 

\begin{acknowledgments}
We thank Bortolo Mognetti for a critical reading of this paper and for several insightful comments that helped us improve it. 
The work of J.F.L. was supported by the European Space Agency (ESA) and the
Belgian Federal Science Policy Office (BELSPO) in the framework of the PRODEX
Programme, Contract No. ESA AO-2004-070. That of C.S. was funded by the
Belgian National Fund for Scientific Research (FRS-FNRS) under the FRIA grant
FC 38825. Computational resources have been provided by the Consortium des Équipements de Calcul Intensif (CÉCI), funded by the Fonds de la Recherche Scientifique de Belgique (F.R.S.-FNRS) under Grant No. 2.5020.11 and by the Walloon Region.
\end{acknowledgments}

\bibliography{mnt.bib}

\appendix

\section{Discretized models}
\label{Discretization}
The points on the computational lattice are labeled $I\equiv I\left(
i_{x},i_{y},i_{z}\right)  $ with $0\leq i_{x}<N_{x}$, etc. and $0\leq I\leq
N_{x}N_{y}N_{z}$. We denote the set of all points as $\mathcal{V}$ (for
volume) and we denote the boundary points as $\partial\mathcal{V\equiv
}\left\{  I\left(  i_{x},i_{y},i_{z}\right)  :i_{x}=0\text{ or }i_{y}=0\text{
or }i_{z}=0\right\}  $. Our calculations always use periodic boundaries so
that, e.g., $I\left(  i_{x}+N_{x},i_{y},i_{z}\right)  =I\left(  i_{x}%
,i_{y},i_{z}\right)  $. If $\overrightarrow{n}=\left(  n_{1},...,n_{N_{x}%
N_{y}N_{z}}\right)  $ be the local density, then the free energy $\beta
\Omega\left(  \overrightarrow{n}\right)  =\beta F\left(  \overrightarrow
{n}\right)  -\beta\mu N\left(  \overrightarrow{n}\right)  $ where $N\left(
\overrightarrow{n}\right)  $ is the total mass, calculated as described in the
main text, $\mu$ is the chemical potential and $F\left(  \overrightarrow
{n}\right)  $ is the Helmholtz functional of cDFT. The latter is written as a
sum of an ideal gas contribution, a hard-sphere contribution and a mean-field
contribution. Our calculations of the free energy use the discretized forms
for these quantities described in Ref.\cite{Schoonen}.

Given the explicit form of the free energy functional, the discretization of
the dynamics only requires specification of the kinetic prefactor, $g^{IJ}%
(n)$. The calculations presented here were performed using
\[
g^{IJ}=\frac{\sigma^2}{\Delta^{5}}\sum_{a=x,y,z}\left\{
\begin{array}
[c]{c}%
{\frac{n^{I+e^{(a)}}+n^{I}}{2}(\delta^{I+e^{(a)}J}-\delta^{IJ})}\\
{-\frac{n^{I}+n^{I-e^{(a)}}}{2}(\delta^{IJ}-\delta^{I-e^{(a)J}})}%
\end{array}
\right\}
\]
where $e^{(a)}$ is the unit vector in the direction $a$ so that e.g.
$I+e^{(x)}$ = $(I_{x}+1,I_{y},I_{z})$. We note that corresponding to this we
have
\begin{equation}
q^{IK(a)}=\frac{\sigma}{\Delta^{5/2}}(\delta^{I+e^{(a)}K}-\delta^{IK})\sqrt{\frac{n^{K}%
+n^{K-e^{(a)}}}{2}}%
\end{equation}
and it is straightforward to confirm that these satisfy the
fluctuation-dissipation relation which, incidentally, assures that $g^{IJ}$ is
positive semi-definite.

\section{Boundary conditions}
\label{BC}
In calculating the free energy and its derivatives, we always use periodic
boundaries (for reasons of efficiency: the calculations are only feasible with
fast Fourier transforms). In order for the dynamics to model an open system,
we take $n^{I}$ to be constant on the border of the computational cell (defined as values of $I$ for
which one or more components is zero). In order that the uniform density be a
stationary solution of the deterministic dynamics, we must also set the forces
$\partial F/\partial n^{I}=\Delta^{3} \mu$ for all values of $I$ on the boundary.
The constant $\mu$ is simply the chemical potential corresponding to the
vapor. Examination of the resulting difference equations shows that a
mathematically equivalent formulation, that is perhaps more physically useful,
is to replace the free energy by $F(n)-\mu N(n)$ and to set $F_{I}=0$ for the
boundary points. This formulation shows that the boundary conditions force the
use of a \textquotedblleft grand canonical\textquotedblright\ free energy
functional. We also note that the fixed density along the borders breaks
translational symmetry (which is otherwise present with periodic boundaries)
and as a result, the matrix $g^{IJ}$ has no zero eigenvalues and is invertible.

To avoid any confusion, we note that the periodic boundaries constitute a \textit{spatial} boundary condition  used to calculated the free energy and its derivatives whereas the fixed density on the borders is a \textit{dynamical} boundary condition pertaining to the solution of the time-dependent equations. The two are therefore independent and compatible as long as the fixed boundary densities are also periodic, as is the case in our calculations.  

\section{Calculation of eigenvalues and eigenvectors}
\label{Eigen}
We determine the eigenvalues and eigenvectors using the SLEPC library using
the Jacobi-Davidson method for the Hessian of the free energy and the
Krylov-Schur method for the dynamical matrix. In both cases,  the Generalized
Minimal Residual method (KSPGMRES) linear equation solver was used. When
diagonalizing the dynamical matrix, we used two-sided balancing and demanded
that the left and right eigenvectors be calculated.

Suppose that we determine a right eigenvector of the dynamical matrix , $v^{I}$ . Then it is easy to see that $v_I \equiv  g_{IJ}v^I$ is a left eigenvector having the same eigenvalue. We can therefore create a normalized, conjugate pair of eigenvectors by defining 
\begin{align}
\widetilde{v}^{I}  &  =\frac{1}{\sqrt{v^{J}g_{JK}v^{K}}}v^{I} \\
\widetilde{v}_{I}  &  =\frac{1}{\sqrt{v^{J}g_{JK}v^{K}}}g^{IL}v_{L}\nonumber
\end{align}
satisfying $\widetilde{v}_{I} = g_{IJ} \widetilde{v}^{J}$  and $\widetilde{v}_{I}\widetilde{v}^{I}=1$. Similarly, given any left eigenvector $u_I$ an analogous pair 
\begin{align}
\widetilde{u}_{I}  &  =\frac{1}{\sqrt{u_{J}g^{JK}u_{K}}}u_{I}\\
\widetilde{u}^{I}  &   =\frac{1}{\sqrt{u_{J}g^{JK}u_{K}}}g^{IL}u_{L}
\nonumber
\end{align}
also exists. All of our results are written with the assumption that the eigenvectors obey this normalization. Finally, if we have independently determined left and right eigenvectors, $u_I$ and $v^I$ , then they should be related by $v^I=\alpha g^{IJ}u_J$ for some constant, $\alpha$, where contraction with $u_I$ gives  
\[
\alpha = \frac{v^Iu_I}{u_Jg^{JK}u_K}
\]
and so $v^{J}g_{JK}v^{K}=\alpha u_kv^K = \frac{\left(v^Iu_I\right)^2}{u_Jg^{JK}u_K}$ giving
\begin{align}
\widetilde{v}^{I}  &  =\frac{\sqrt{u_{J}g^{JK}u_{K}}}{u_Lv^L}v^{I} \\
\widetilde{v}_{I}  &  =\frac{1}{\sqrt{u^{J}g^{JK}u_{K}}}u_{I}\nonumber
\end{align}

\section{Calculation details}
\label{Details}
The spherically symmetric pair-potentials used in this study were a cut-off
Lennard-Jones (LJ) potential,
\begin{equation}
v_{\text{LJ}}\left(  r\right)  =4\varepsilon\left(  \left(  \frac{\sigma}
{r}\right)  ^{12}-\left(  \frac{\sigma}{r}\right)  ^{6}\right)  -4\varepsilon
\left(  \left(  \frac{\sigma}{r_{\text{cut}}}\right)  ^{12}-\left(
\frac{\sigma}{r_{\text{cut}}}\right)  ^{6}\right)  ,\;r<r_{\text{cut}}%
\end{equation}
with cutoff $r_{\text{cut}}=3\sigma$ and the potential of Wang et
al\cite{Wang} or WHDF potential,%
\begin{equation}
v_{\text{WHDF}}\left(  r\right)  =114.11\epsilon\left(  \left(
\frac{\sigma}{r}\right)  ^{2}-1\right)  \left(  \left(  \frac{r_{\text{cut}}%
}{r}\right)  ^{2}-1\right)  ^{2},\;r<r_{\text{cut}}%
\end{equation}
with cutoff $r_{\text{cut}}=1.2\sigma$. The computational lattice spacing was
$\Delta=0.2\sigma$ for the Lennard-Jones and $\Delta=0.1\sigma$ for the WHDF potential. Both potentials have a nearly hard-core repulsion at short distances ($r < \sigma$) and an attractive well outside this. The difference between them is that the ratio of the width of the attractive part of the potential to the width of the hard-core region is much smaller for WHDF than for the LJ. This is intended to model typical colloidal interactions.

To find critical clusters, we first minimize the free energy of a system consisting of a droplet in a background of low-density fluid under fully periodic boundary conditions (e.g. no fixed density at the boundary): this implies  constant total number of particles and so a canonical-like minimization. It has previously been shown that such a cluster is a stationary point when the system is opened: that is, when the density is held constant at the boundary and the total number of particles allowed to vary\cite{LutskoLam,LutskoHCF}. In this grand-canonical system, the cluster may be either a local minimum or a saddle point - there is no way to know except to test it, e.g. by calculating the eigenvalues. Notice that when this is done at sufficiently low temperatures, the clusters spontaneously form solid-like, amorphous or crystalline, order\cite{LutskoLam}. 

Regardless of the nature of the cluster so obtained, we use it as the endpoint for the determination of the nucleation pathway using the string method as described in detail in Ref.\cite{LutskoHCF}. If the cluster anchoring the string was a local minimum, a critical cluster appears somewhere along this pathway. However, because of the limited resolution of the string method - i.e. the fact that the pathway is characterized by a finite number of images - the "critical cluster" identified is not in fact exactly a saddle point of the free energy surface. We therefore further refine it using various methods (e.g. an eigenvector-following method\cite{Wales}) to be described in a future publication. This refined critical cluster is then used as the endpoint for a new string calculation and, in the case of LJ reported in the main text, a separate string calculation was used to generate the solid-solid pathway from the refined critical cluster to the endpoint of the initial string calculation. 

\section{Measure of order within the clusters} \label{order}
We track the development of order in the clusters using a quantity that we
call "localization". It is ignorant of the specific type of solid structure
that is appearing and simply measures how much the density is peaked around
the "atomic" positions. We define it in terms of the local packing fraction
$\eta^{I}(R)$, which measures the amount of mass within a radius $R$ around a
lattice position $I$. The evaluation of these packing fractions on the
computational lattice is not trivial and is explained in \cite{Schoonen}. The
radius $R$ is the effective hard sphere radius of our cDFT model. We estimate
how much the density is localized around a position $I$ by computing the
excess of mass $\eta^{I}(R/2) - \eta^{I}(R)/8$ within a smaller radius $R/2$,
which we then normalize to get
\begin{equation}
l^{I} = \frac{\eta^{I}(R/2) - \eta^{I}(R)/8}{\eta^{I}(R)-\eta^{I}(R)/8}%
\end{equation}
This (local) measure of localization is zero if the mass $\eta^{I}(R)$ is
uniformly distributed in the sphere of radius $R$ and one if this entire mass
is located in the smaller sphere. The density-weighted average of $l^{I}$ over
the entire computational lattice is our (global) measure of localization
within the cluster,
\begin{equation}
\bar l = \frac{1}{N} \sum_{I} n^{I} l^{I} \Delta^{3}%
\end{equation}
where $N=\sum_{I} n^{I} \Delta^{3}$ is the total number of particles.

\section{Calculating distance along the string path}
\label{string_details}
The string method results in a series of snapshots of the system along the
nucleation pathway. We label these as $n_{\alpha}$ with $\alpha= 0$
corresponding the initial, uniform system and $\alpha= 31$ corresponding to
the final critical clusters shown in the Figures. We approximate the distance
between successive points along the pathway by defining the difference $\Delta
n^{I}_{\alpha+1/2} = n^{I}_{\alpha+1}-n^{I}_{\alpha}$ and the average $\bar
{n}^{I}_{\alpha+1/2} = \frac{1}{2}\left( n^{I}_{\alpha+1}+n^{I}_{\alpha
}\right) $ and evaluating
\begin{equation}
\Delta s(\alpha,\alpha+1) = \sqrt{\Delta n^{I}_{\alpha+1/2} g_{IJ}\left(
\bar{n}_{\alpha+1/2}\right)  \Delta n^{J}_{\alpha+1/2}}%
\end{equation}
and then using $s(\alpha) = \sum_{\gamma=1}^{\alpha}\Delta s(\gamma-1,\gamma)$.
Since we only have an explicit representation for the matrix $g^{IJ}$, and we
require the inverse for this evaluation, we write
\begin{equation}
g^{IJ}\phi_{J}= \Delta n^{I}_{\alpha+1/2}%
\end{equation}
and solve this system of linear equations for $\phi_{I}$ using tools from the
PETSC library\cite{petsc-user-ref,petsc-efficient}. We can then replace
$\Delta n^{I} g_{IJ} \Delta n^{J}$ by $\Delta n^{I} \phi_{I}$.

\section{Using MeNT to determine the CNT nucleation rate}
\label{Rates}
The CNT nucleation rates require as input the free energy at the critical cluster, the second derivative of the free energy with respect to particle number at the critical cluster and the attachment rate. The free energies are of course directly given from our model and the derivatives of the free energy were discussed in a previous section. 
 To determine the CNT attachment and detachment frequencies (which are only needed at the critical cluster in order to calculate the CNT nucleation rates) we compare the stochastic model near the critical cluster to the assumed dynamics of CNT. To this end, the stochastic equation is expanded to first order to get
\begin{equation}
\frac{d}{dt}\delta\widehat{n}_{t}^{I}=-\frac{D}{\sigma^{2}}g^{IJ}\left(  n^{\ast
}\right)  \left(  \frac{\partial^{2}\beta\Omega\left(  n\right)  }{\partial
n^{J}\partial n^{K}}\right)  _{n^{\ast}}\delta\widehat{n}_{t}^{K}+\sqrt
{\frac{2D}{\sigma^{2}}}{\mathbf q}^{IJ}\left(  n^{\ast}\right) \cdot \widehat{\boldsymbol{\xi}%
}_{t}^{J}.
\end{equation}
Using the fact that $\delta N=M_{I}\delta\widehat{n}^{I}\simeq\left(
M_{J}v^{\left(  -\right)  J}\right)  v_{I}^{\left(  -\right)  }\delta
\widehat{n}^{I}$ we find%
\begin{align} \label{NearCC}
\frac{d}{dt}\delta N &  =-\frac{D}{\sigma^{2}}\left(  M_{L}v^{\left(  -\right)
L}\right)  v_{I}^{\left(  -\right)  }g^{IJ}\left(  n^{\ast}\right)  \left(
\frac{\partial^{2}\beta\Omega\left(  n\right)  }{\partial n^{J}\partial n^{K}%
}\right)  _{n^{\ast}}\delta\widehat{n}_{t}^{K}\\
&  \;\;\;\;\;\;\;\;\;\;+\sqrt{\frac{2D}{\sigma^{2}}}\left(  M_{L}v^{\left(
-\right)  L}\right)  v_{I}^{\left(  -\right)  }{\mathbf q}^{IJ}\left(  n^{\ast
}\right)  \cdot \widehat{\boldsymbol{\xi}}_{t}^{J}\nonumber\\
&  =-\frac{D}{\sigma^{2}}\lambda^{\left(  -\right)  }\left(  M_{L}v^{\left(
-\right)  L}\right)  v_{I}^{\left(  -\right)  }\delta\widehat{n}_{t}%
^{I}\nonumber\\
&  \;\;\;\;\;\;\;\;\;\;\;+\sqrt{\frac{2D}{\sigma^{2}}}\left(  M_{L}v^{\left(
-\right)  L}\right)  v_{I}^{\left(  -\right)  }{\mathbf q}^{IJ}\left(  n^{\ast
}\right)  \cdot \widehat{\boldsymbol{\xi}}_{t}^{J}\nonumber\\
&  \equiv -\frac{D}{\sigma^{2}}\lambda^{\left(  -\right)  }\delta N+\sqrt{\frac{2D}{\sigma^{2}%
}}\left(  M_{L}v^{\left(  -\right)  L}\right)  \widehat{\xi}%
_{t}\nonumber
\end{align}
where, in the last step, we have replaced the white noise by a simpler form
having the same covariance and therefore corresponding to the same Fokker-Planck equation\cite{Gardiner}.

In CNT, it is assumed that the size of clusters changes by the attachment and
detachment of monomers and that these occur with rates $f\left(  N\right)  $
and $g\left(  N\right)  $ for a cluster of size $N$. This means that the
probability that a cluster has size $N$ at time $t+\tau$ is, for small $\tau
$,
\begin{align}
P\left(  N;t+\tau\right)    & =P\left(  N;t\right)  +\tau f\left(  N-1\right)
P\left(  N-1;t\right)  \\
& -\tau\left(  f\left(  N\right)  +g\left(  N\right)  \right) P(N;t) +\tau g\left(
N+1\right)  P\left(  N+1;t\right)  \nonumber
\end{align}
which can be rearranged as
\begin{align}
\frac{P\left(  N;t+\tau\right)-P\left(  N;t\right)}{\tau}  & = \frac{\delta_2\left(\left(g(N)-f(N)\right)P(N;t)\right)}{2} \\
&+ \frac{\delta^2_1\left(\left(f(N)+g(N)\right)P(N;t)\right)}{2} \nonumber
\end{align}
where we have used the standard notation for centered finite differences, viz. $\delta_hf(x) \equiv f(x+\frac{h}{2})-f(x-\frac{h}{2})$.
Taking the limit that $\tau\rightarrow0$ leads to a differential equation in time, treating $N$ as a continuous variable and replacing the finite differences by derivatives gives a partial differential equation
for $P\left(  N;t\right)  $ that is first order in time and second order in
$N$ ,%
\begin{equation} \label{FP1}
\frac{\partial}{\partial t}P\left(  N,t\right)  =\frac{\partial}{\partial
N}\left(
\begin{array}
[c]{c}%
\left(  g\left(  N\right)  -f\left(  N\right)  \right)  P\left(
N,t\right)  \\
+\frac{1}{2}\frac{\partial}{\partial N}\left[\left(  f\left(  N\right)  +g\left(
N\right)  \right)  P\left(  N,t\right)\right]
\end{array}
\right)
\end{equation}
and is recognized as a Fokker-Planck equation. (This is essentially the same
as the Tunitskii equation of CNT\cite{Kashchiev}). That Fokker-Planck equation
is in turn equivalent to the (Ito-) stochastic equation\cite{Gardiner} 
\begin{equation}
\frac{d}{dt}\widehat{N}_{t}= \left(f\left(  \widehat{N}_{t}\right)  -g\left(
\widehat{N}_{t}\right)\right)   +\sqrt{ f\left(  \widehat{N}_{t}\right) + g\left(  \widehat{N}_{t}\right)}  \widehat{\mathbf{\xi}}_{t}
\label{stoch}%
\end{equation}
Comparing this to Eq.(\ref{NearCC}) and evaluating both at the critical cluster,  we can identify the
attachment and detachment frequencies as 
\begin{equation} \label{frequencies}
f(N^{\ast})=g(N^{\ast})=\frac{D}{\sigma^{2}}\left(  M_{J}v^{\left( -\right)
J}\right)  ^{2} = \frac{D}{\sigma^{2}}\left(\frac{dN}{ds}\right)_{n^\ast}^2.
\end{equation}


To relate these quantities to the thermodynamics, as in CNT, we turn to the evaluation of 
the derivative of the free energy along the MLP. The derivative of the free energy with respect to particle 
number along the MLP is evaluated using the fact that the MLP points along the deterministic
driving force (in the weak noise approximation) so near the given point on the
MLP, $n^{(\text{MLP})}$, the path can be parameterized as
\begin{equation}
n^{(\text{MLP})I}\left(  \alpha\right)  =n^{(\text{MLP})I}+\alpha g^{IJ}%
\beta\Omega_{J}(n^{(\text{MLP})})
\end{equation}
and so
\begin{align}
\left.  \frac{d\beta\Omega}{dN}\right\vert _{n^{(\text{MLP})}}  &
=\lim_{\alpha\rightarrow0}\left(  \frac{d\beta\Omega\left(  n^{(\text{MLP}%
)}\left(  \alpha\right)  \right)  }{d\alpha}\right)  /\left(  \frac{dN\left(
n^{(\text{MLP})}\left(  \alpha\right)  \right)  }{d\alpha}\right)  \\
& =\frac{\beta\Omega_{I}\left(  n^{(\text{MLP})}\right)  g^{IJ}\left(
n^{(\text{MLP})}\right)  \beta\Omega_{J}\left(  n^{(\text{MLP})}\right)
}{M_{K}g^{KL}\left(  n^{(\text{MLP})}\right)  \beta\Omega_{L}\left(
n^{(\text{MLP})}\right)  }.\nonumber
\end{align}
At the critical point, the first derivative of the free energy vanishes and so, near the critical point, the steepest descent equation for $\delta
n^{I}=n^{I}-n^{\ast I}$ takes the form%
\begin{equation}
\frac{d\delta n^{I}}{dt}=\frac{D}{\sigma^2}g^{IJ}\left(  n^{\ast}\right)  \beta\Omega
_{JK}\left(  n^{\ast}\right)  \delta n^{K}%
\end{equation}
and the displacement along the MLP obeys
\begin{equation}
\delta n^{I}\propto g^{IJ}\left(  n^{\ast}\right)  \beta\Omega_{JK}\left(
n^{\ast}\right)  \delta n^{K},
\end{equation}
or in other words, it is the direction of the unstable eigenvector. Hence, we
now have
\begin{equation}
n^{(\text{MLP})I}\left(  \alpha\right)  =n^{\ast I}+\alpha v^{\left(
-\right)  I}%
\end{equation}
and straightforward evaluation gives
\begin{equation} \label{Derivatives}
\left.  \frac{d\beta\Omega}{dN}\right\vert _{n^{\ast}}=0,\;\left.  \frac
{d^{2}\beta\Omega}{dN^{2}}\right\vert _{n^{\ast}}=\lambda^{\left(  -\right)
}\frac{v_{J}^{\left(  -\right)  }v^{\left(  -\right)  J}}{\left(
M_{I}v^{\left(  -\right)  I}\right)  ^{2}}= \frac{\lambda^{\left(  -\right)  }%
}{\left(  M_{I}v^{\left(  -\right)  I}\right)  ^{2}}.
\end{equation}

It is also useful below to note that along the MLP,
\begin{equation}
    ds=\sqrt{dn^Ig_{IJ}dn^J}=\sqrt{\beta \Omega_Ig^{IJ}d\beta\Omega_J}d\alpha
\end{equation}
and so
\begin{equation}
    \frac{dN}{ds}=\frac{M_Kg^{KL}\beta \Omega_L}{\sqrt{\beta\Omega_Ig^{IJ}d\beta\Omega_J}}.
\end{equation}
At the critical cluster, these become $ds = d\alpha$ and 
\begin{equation}
    \left(\frac{dN}{ds}\right)_{n^\ast}=M_Iv^{(-)I},
\end{equation}
respectively.

Finally, using Eq.(\ref{frequencies}) and (\ref{Derivatives}), to evaluate the CNT expression for the nucleation rate, Eq.(\ref{JCNT}) gives  Eq.(\ref{rate}).

\section{Heuristic one-dimensional model}
\label{Heuristic}
To give an idea of the effect of the entire nucleation pathway on the
nucleation rates, we consider a 1-D model based on the information derived
from the full theory. In the main text, we used a projection of the stochastic
equation onto the unstable mode at the critical cluster to extract the
attachment rate. Here, we follow a similar idea and, as in the discussion
above (Derivatives along the path) we project in the direction of the
deterministic force giving
\begin{align}
\frac{\beta\Omega_{I}}{\sqrt{\beta\Omega_{I}g^{IJ}\beta\Omega_{J}}}\frac{d}
{dt}\widehat{n}_{t}^{I}  & =-\frac{D}{\sigma^{2}}\frac{\beta\Omega_{I}}{\sqrt
{\beta\Omega_{I}g^{IJ}\beta\Omega_{J}}}g^{IJ}\beta\Omega_{J}\\
& +\sqrt{\frac{2D}{\sigma^{2}}}\frac{\beta\Omega_{I}}{\sqrt{\beta\Omega_{I}g^{IJ}%
\beta\Omega_{J}}}{\mathbf q}^{IJ} \cdot \widehat{\boldsymbol{\xi}}_{t}^{J},\nonumber
\end{align}
where we use a short-hand notation in which $\beta\Omega_{I}\equiv
\partial\beta\Omega/\partial n^{I}$. To make sense of this, consider the
deterministic evolution of the distance which obeys%
\begin{equation}
\frac{ds}{dt}=\sqrt{ \frac{dn^{I}}{dt}g_{IJ}\frac{dn^{J}}{dt}}=\frac{D}{\sigma^{2}%
}\sqrt{\beta\Omega_{I}g^{IJ}\beta\Omega_{J}}%
\end{equation}
so%
\begin{equation}
\frac{\beta\Omega_{I}}{\sqrt{\beta\Omega_{I}g^{IJ}\beta\Omega_{J}}}\frac{dn^{I}%
}{dt}=-\frac{ds}{dt}.%
\end{equation}
This suggests, then, the model (Stratonovich\cite{Gardiner}) stochastic differential equation
\begin{align}
\frac{ds}{dt}  & =-\frac{D}{\sigma^{2}}\sqrt{ \beta\Omega_{I}g^{IJ}\beta\Omega_{J}%
}+\sqrt{\frac{2D}{\sigma^{2}}}\frac{\beta\Omega_{I}}{\sqrt{\beta\Omega
_{I}g^{IJ}\beta\Omega_{J}}}{\mathbf q}^{IJ} \cdot \widehat{\boldsymbol{\xi}}_{t}^{J},
\end{align}
and, again replacing the noise term by an equivalent one with the same covariance,
this becomes%
\begin{equation}
\frac{ds}{dt}=-\frac{D}{\sigma^{2}}\sqrt{\beta\Omega_{I}g^{IJ}%
\beta\Omega_{J}}+\sqrt{\frac{2D}{\sigma^{2}}}\widehat{\mathbf{\xi}%
}_{t}.%
\end{equation}
Finally, since all motion is along the MLP, $\sqrt{\beta\Omega_{I}g^{IJ}\beta\Omega_{J}}=\frac{d\beta\Omega}{ds}$,
\begin{equation} \label{Nt}
\frac{ds}{dt}=-\frac{ D}{\sigma^{2}}\frac
{d\beta\Omega\left(s\right)  }{ds}+\sqrt{\frac{2D}{\sigma^{2}}}\widehat{\mathbf{\xi}}_{t}.%
\end{equation}
We pause to note that the equilibrium distribution for this stochastic process will be $P(s) \propto e^{-\beta \Omega(s)}$ as one would expect if the variable $s$ is the correct order parameter. 

A detailed analysis of this model will be given elsewhere. Here, we note that the mean first passage time, $T(s)$, for a cluster in state $s$ to eventually reach the critical state, $s^\ast$, corresponding to the critical cluster can be determined exactly\cite{Gardiner} and from this, the nucleation rate is given by the population over the flux\cite{Hanggi} giving the general expression
\begin{equation}
J\simeq\frac{D}{\sigma^{2}}\frac{1}{2}\int_{s_0}^{s^\ast} \frac{c(s)}{\int_{s}^{s^{\ast}}e^{\beta\Omega\left(  s'\right)
}\left(\int_0^{s'}e^{-\beta\Omega(s'')}ds''\right)ds'}ds\label{J1}%
\end{equation}
where $c_{1}$ is the concentration of monomers . If we assume that the system begins in a (near-) equilibrium state, $c(N) \approx c_1e^{-\beta \left(\Omega(N)-\Omega(1)\right)}$, then $c(s) \approx c_1e^{-\beta \left(\Omega(s)-\Omega(0)\right)}\frac{dN}{ds}$ and this can be evaluated. (Note that the factor of $\frac{dN}{ds}$ is necessary as $c(N)$ and $c(s)$ are densities. For example, integrating $c(N)$ over $N=0$ to $N^\ast$ has to be the same as integrating $c(s)$ over $s=0$ to $s^\ast$ since both are the total number of sub-critical clusters.) Standard approximations give the intermediate form
\begin{equation}
J\simeq\frac{D}{\sigma^{2}}\frac{1}{2} \left(\frac{dN}{ds}\right)_{s=0}\frac{c_1}{\int_{0}^{s^{\ast}}e^{\beta\Delta \Omega\left(  s\right)
}ds}\label{J2}%
\end{equation} 
and a final evaluation using Laplace's method gives 
\begin{equation}
    J \simeq \frac{D}{\sigma^{2}} c_1 \left(\frac{dN}{ds}\right)_{s=0} \sqrt{\frac{1}{2\pi} \left| \frac{d^2}{ds^2}\beta\Omega(s) \right|_{s=s^\ast} }e^{-\beta \Delta \Omega^\ast}
\end{equation}
or, changing variables, 
\begin{equation}
    J \simeq \frac{D}{\sigma^{2}} c_1 \left(\frac{dN}{ds}\right)_{N=1} \left(\frac{dN}{ds}\right)_{N^\ast}\sqrt{\frac{1}{2\pi} \left| \frac{d^2}{dN^2}\beta\Omega(N) \right|_{N=N^\ast} }
\end{equation}
Not surprisingly, this only agrees with the CNT result, given in Eq.(\ref{rate}) of the main text, if $\left(\frac{dN}{ds}\right)_{N=1} = \left(\frac{dN}{ds}\right)_{N\ast}$, or, more generally, if $\frac{dN}{ds}$ is constant, which means that $s$ and $N$ are equivalent as order parameters. 

It is interesting to note that we can use this model to make identify the attachment and detachment frequencies for any sized cluster. First, we multiply Eq.(\ref{Nt}) by $dN/ds$ go get the Stratonovich stochastic differential equation
\begin{equation} \label{Nt2}
\frac{dN}{dt}=-\frac{D}{\sigma^{2}}\left(\frac{dN}{ds}\right)^2\frac
{d\beta\Omega\left(N\right)  }{ds}+\sqrt{\frac{2D}{\sigma^{2}}}\frac{dN}{ds}\widehat{\mathbf{\xi}}_{t}.%
\end{equation}
and which is equivalent to the Fokker-Planck equation
\begin{equation}
\frac{\partial}{\partial t}P(N,t)=\frac{D}{\sigma^2}\frac{\partial}{\partial N} 
\left(\left(\frac{dN}{ds}\right)^2\frac{d\beta\Omega(N)}{dN}+\frac{dN}{ds} \frac{\partial}{\partial N} \frac{dN}{ds}
\right) P(N,t)
\end{equation}
which can also be written as
\begin{equation}
\frac{\partial}{\partial t}P(N,t)=\frac{D}{\sigma^2}\frac{\partial}{\partial N} 
\left(
\begin{array}[c]{c}
\left(\frac{dN}{ds}\right)^2\frac{d\beta\Omega(N)}{dN}-\frac{dN}{ds}\frac{d}{dN}\frac{dN}{ds} \\
+ \frac{\partial}{\partial N} \left(\frac{dN}{ds}\right)^2
\end{array}
\right) P(N,t)
\end{equation}
or more succinctly,
\begin{equation} \label{FPN}
\frac{\partial}{\partial t}P(N,t)=\frac{D}{\sigma^2}\frac{\partial}{\partial N} 
\left(
\begin{array}[c]{c}
  \left(\frac{dN}{ds}\right)^2\frac{d\left(\beta\Omega(N)-\ln\frac{dN}{ds}\right)}{dN}\\ + \frac{\partial}{\partial N} \left(\frac{dN}{ds}\right)^2
\end{array}
  \right) P(N,t)
\end{equation}
Comparing to the CNT result, Eq.(\ref{FP1}), we can identify
\begin{align}
f\left(  N\right)  -g\left(  N\right)   &  = - \frac{D}{\sigma^2} \left(  \frac{dN}{ds}\right)
^{2}\frac{d\left(\beta\Omega\left(  N\right)  +\ln\frac{ds}{dN}  \right)}{dN}\\
f\left(  N\right)  +g\left(  N\right)   &  =2\frac{D}{\sigma^2}\left(  \frac{dN}{ds}\right)
^{2}.\nonumber
\end{align}
At the critical cluster, by definition $f(N^{*}) = g(N^{*})$ so this gives $f(N^{*}) = \frac{D}{\sigma^2}\left(\frac{dN}{ds}\right)_{N\ast} = \frac{D}{\sigma^2}\left(M_{I}v^{(-)I}\right)^2$ as in the main text. Note however that the critical cluster is not what one might expect: it is not the derivative of the free energy that is zero but of the shifted free energy $\beta \Omega + \frac{ds}{dN}$. This is related to an assumption that permeates CNT although it is seldom discussed. To illustrate within the present context, let us imagine a sub-critical - and therefore, equilibrium - fluid. In this case, the probability that a cluster has a given size should be constant and in fact it is easy to determine the equilibrium distribution from our model. Referring to Eq. (\ref{FP1}), one easily finds that the probability that a cluster has size $N$, $P(N)$,  is $P(N) = \mathcal{N} e^{-\left(\beta \Omega(N) + \ln \frac{ds}{dN}\right)}$, where the prefactor is a normalization constant. Hence, the effective free energy is shifted. The origin of this shift is apparent if we change variables from the size of the cluster to its order parameter. Assuming an invertible relation, $s(N)$, the probability that a cluster has order parameter $s$, $\tilde{P}(s) = P(N(s))\frac{dN}{ds}$  where the second factor on the right is needed to preserve normalization (integrating over all $s$ must be the same as integrating over all $N$). Working this out gives $\tilde{P}(s) = \mathcal{N}e^{-\beta \tilde{\Omega}(s)}$ where $\tilde{\Omega}(s) \equiv \Omega(N(s))$. So, the probability has the canonical form when written in terms of the natural order parameter and it is only when making a nonlinear change of variable that the "shift" occurs. In CNT, one sometimes sees presentations in which the order parameter is not taken to be $N$ but, rather, the radius of a cluster $R$. In that case, it is often assumed that an equilibrium distribution is proportional to $e^{-\beta \Omega(R)}$ rather than $e^{-\beta \Omega(N)}$ and it is clear that both cannot be true. The fact that the free energy is shifted in these expressions reflects the fundamental observation that the "classical" picture really only fully holds when $s$ and $N$ are linearly related: curvature in their relation indicates that changes in the local density occur which are not directly related to changes in total mass and this violates the fundamental assumptions of CNT.  We will explore these issues more fully in a future publication.

\end{document}